\documentclass[floatfix,twocolumn,aps,showpacs,superscriptaddress]{revtex4}

\usepackage{amsmath,amssymb}
\usepackage[]{graphicx}       
\usepackage{dcolumn}          
\usepackage{bm}               
\usepackage[dvips]{color}     

\input{MyMc}

\newcommand{\av}[2][]{\ensuremath{\left\langle#2\right\rangle}_{#1}\xspace}

\newcommand{\mean}[1]{\ensuremath{\overline{#1}}\xspace}

\newcommand{\Idc}{\ensuremath{I_\mathrm{ dc}}\xspace}
\newcommand{\Iac}{\ensuremath{I_\mathrm{ ac}}\xspace}
\newcommand{\Icp}{\ensuremath{I_{c+}}\xspace}
\newcommand{\Icn}{\ensuremath{I_{c-}}\xspace}
\newcommand{\Istop}{\ensuremath{I_\mathrm{stop}}\xspace}
\newcommand{\Isupp}{\ensuremath{I_\mathrm{supp}}\xspace}

\begin{document}

\title{%
  Deterministic Josephson Vortex Ratchet with a load
}

\author{M.~Knufinke}
\affiliation{%
  Physikalisches Institut -- Experimentalphysik II and Center for Collective Quantum Phenomena in LISA$^+$,
  Universit\"{a}t T\"{u}bingen,
  Auf der Morgenstelle 14,
  D-72076 T\"{u}bingen, Germany
}%

\author{K.~Ilin}
\author{M.~Siegel}
\affiliation{%
  Universit\"{a}t Karlsruhe,
  Institut f\"{u}r Mikro- und Nanoelektronische Systeme,
  Hertzstra\ss e 16,
  D-76187 Karlsruhe, Germany
}%

\author{D.~Koelle}%
\author{R.~Kleiner}%
\author{E.~Goldobin}
\affiliation{%
  Physikalisches Institut -- Experimentalphysik II and Center for Collective Quantum Phenomena in LISA$^+$,
  Universit\"{a}t T\"{u}bingen,
  Auf der Morgenstelle 14,
  D-72076 T\"{u}bingen, Germany
}%


\begin{abstract}

We investigate experimentally a deterministic underdamped Josephson vortex ratchet -- a fluxon-particle moving along a Josephson junction in an asymmetric periodic potential. By applying a sinusoidal driving current one can compel the vortex to move in a certain direction, producing average dc voltage across the junction. Being in such a rectification regime we also load the ratchet, i.e., apply an additional dc bias current $I_\mathrm{dc}$ (counterforce) which tilts the potential so that the fluxon climbs uphill due to the ratchet effect. The value of the bias current at which the fluxon stops climbing up defines the strength of the ratchet effect and is determined experimentally. This allows us to estimate the loading capability of the ratchet, the output power and efficiency. For the quasi-static regime we present a simple model which delivers simple analytic expressions for the above mentioned figures of merit.

\end{abstract}


\date{\today}

\maketitle

\section{Introduction}
\label{sec:Introduction}

The discovery of Brownian motion gave birth to the idea of extracting useful work out of random motion. As Richard Feynman \etal demonstrated \cite{Feynman1966}, drawing energy from equilibrium thermal fluctuations (white noise) is forbidden by the second law of thermodynamics. The extraction of work out of non-equilibrium or time-correlated noise (colored noise) is possible in the so-called ratchet systems. Such systems incorporate an asymmetric periodic potential \cite{Magnasco1993} and have been in the focus of investigation during the last two decades in various implementations. Ratchets based on Josephson junctions, in particular exploiting the motion of the Josephson phase in SQUIDs or vortices in long Josephson junctions (LJJs) have been suggested and tested experimentally \cite{Juelicher1997, Reimann02, Haenggi1996, Haenggi2009, Sterck2002, Sterck2005a, Sterck2009, Weiss2000, Goldobin2001, Carapella2001a, Carapella2001, Carapella02b, Beck2005, Zapata1996}.

These systems have some advantages over other systems: (I) directed motion results in an average dc voltage which makes ratchet operation easily accessible in experiment; (II) Josephson junctions are very fast devices, \ie, they can be operated in a broad frequency range from dc up to $100\,\mathrm{GHz}$ which allows them to capture a lot of spectral energy; (III) both underdamped and overdamped systems can be investigated by proper junction design and the variation of the bath temperature.

The deterministic underdamped Josephson vortex ratchet (JVR) in which a Josephson vortex (fluxon) moves along a LJJ was implemented by our group earlier \cite{Beck2005}. A strongly asymmetric tunable potential was created using a current injector \cite{Beck2005, Goldobin2001}. The periodicity of the potential is provided by the annular geometry of the LJJ\cite{Davidson1985, Davidson1986, Ustinov92, Martucciello96x}. The fluxon was injected into the annular LJJ also in a controllable way using a pair of tiny current injectors \cite{Ustinov02a, Malomed04}. The directional motion of a fluxon was detected by measuring the (averaged) dc voltage across the junction which is, due to the Josephson relation, proportional to the average velocity of a fluxon and was reaching values as high as $0.9{\bar c}_{0}$, where ${\bar c}_{0}$ is the Swihart velocity (maximum possible velocity of a fluxon). However, in these experiments the JVR was idle, i.e. was not delivering any rectified power to a load.

In this paper we investigate several figures of merit of such a JVR relevant for applications: rectification window, maximum dc counterforce against which the ratchet can still work, output power and efficiency in both quasi-static and non-adiabatic regimes.

The paper is organized as follows: In Sec.~\ref{sec:Theory} we describe the equations for the dynamics of the Josephson phase in a LJJ with a ratchet potential created by an injector current. The sample design and characterization are presented in Sec.~\ref{sec:Samples}. Experimental and analytical results for quasi-static driving frequencies are presented in Sec.~\ref{sec:QuasiStaticDrive}. The Sec.~\ref{sec:NonAdiabaticDrive} covers experimental and numerical results for the non-adiabatic drive. Sec.~\ref{sec:Conclusion} concludes this work.

\section{Theory}
\label{sec:Theory}

\begin{figure}[t]
  \begin{center}
    \includegraphics{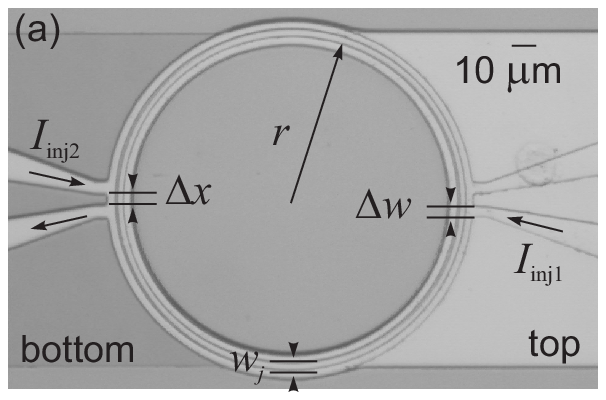}
    \includegraphics{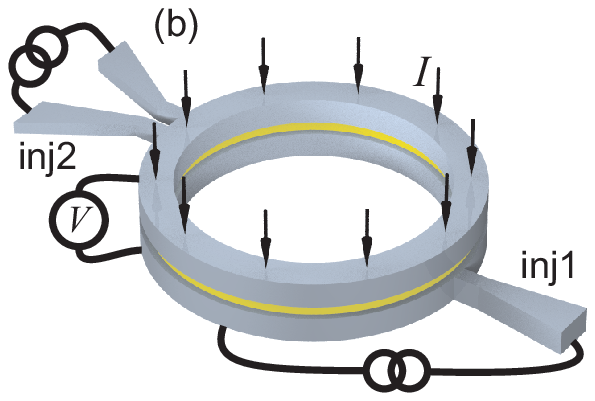}
  \end{center}
  \caption{(Color online)
    (a) Optical image of the ALJJ used in our experiments. Only one injector of the right pair is used.
    (b) Sketch of the ALJJ with a single injector 1 and a double injector 2. Bias leads for top and bottom electrode are not shown for clarity.
  }
  \label{fig:JJ_Layout_PovRay_bb}
\end{figure}

Our system consists of an annular long Josephson junction (ALJJ) equipped with injectors to create an asymmetric potential and to insert a fluxon, see Fig.~\ref{fig:JJ_Layout_PovRay_bb}. The dynamics of the Josephson phase in the system can be described by the following perturbed sine-Gordon equation \cite{Goldobin2001}:
\begin{equation}
  \phi_{xx}-\phi_{tt}-\sin\phi = \alpha\phi_t-\gamma-\gamma_\mathrm{inj}(x)-\xi(t),
  \label{eqn:sine-gordon-equation}
\end{equation}
where $\phi(x,t)$ is the Josephson phase and subscripts $x$ and $t$ denote derivatives with respect of space and time, respectively. The curvilinear coordinate $x$ along the LJJ is normalized to the Josephson penetration depth $\lambda_J$
and the time $t$ is normalized to the inverse of the plasma frequency $\omega_p^{-1}$.
The quantity $\alpha$
is the dimensionless damping parameter. $\gamma=j/j_c$,  $\gamma_\mathrm{inj}(x)=j_\mathrm{inj}(x)/j_c$ and $\xi(t)=j(t)/j_c$ are the dc bias current density, injector current density and ac driving current density, respectively, all normalized to the critical current density $j_c$ of the LJJ. $\gamma_\mathrm{inj}(x)$ has zero spatial average and is used to create an asymmetric potential, $\gamma$ is used to apply an additional dc bias current while the device is operated in the rectification regime (see below). $\xi(t)$ is a spatially homogeneous deterministic (or stochastic) drive with zero time average. The ultimate aim of ratchet operation is to rectify $\xi(t)$ to produce a nonzero voltage $\mean{V}\propto\mean{\phi_t}\neq 0$, which is independent on $x$.

In the absence of the right-hand side, the solitonic solution of Eq.~\eqref{eqn:sine-gordon-equation} is a Josephson vortex (sine-Gordon kink) $\phi(x)=4\arctan\left[\exp\left\{\left[x-x_0(t)\right]/\sqrt{1-u^2}\right\}\right]$ situated at $x_0$ and moving with velocity $u=\mathrm{d}x_0(t)/\mathrm{d}t$ \cite{Ustinov98}. The right-hand side of Eq.~\eqref{eqn:sine-gordon-equation} is usually considered as a perturbation \cite{McLaughlin78}. It does not change drastically the vortex shape, but defines its dynamics, \eg, its equilibrium velocity \cite{McLaughlin78}. Such an approximation essentially treats the vortex as a rigid object, and its dynamics can be reduced to the dynamics of a relativistic underdamped point-like particle \cite{Carapella2001a} (\cf the nonrelativistic case \cite{Borromeo02}). In these terms, the ratchet should rectify $\xi(t)$ to produce a nonzero average velocity $\mean{u}\neq 0$.

We implement the asymmetric potential using a single injector (further called inj1), see Fig.~\ref{fig:JJ_Layout_PovRay_bb}. A suitable current profile $\gamma_\mathrm{inj}(x)$ is equivalent to a non-uniform magnetic field $h(x)$ along the LJJ, such that $\gamma_\mathrm{inj}(x)=-h_x(x)$. The normalized potential $U(x_0)$ felt by a fluxon-particle is given by $U(x_0)\approx -2\pi\left(w_J/\lambda_J\right)h(x_0)$, where $w_J$ is the junction width \cite{Goldobin2001}.

Such a profile can be realized by applying a current $\gamma_\mathrm{inj}$ through a single injector of width $\Delta w$ situated at $x=x_\mathrm{inj1}$ and extracting the current over the same electrode along the rest of the LJJ, \ie,
\begin{equation}
  \gamma_\mathrm{inj}(x)=\left\{
    \begin{array}{ll}
       \gamma_1 & \text{ for } \left|x-x_\mathrm{inj1}\right|<\Delta w/(2\lambda_J)\\
       \gamma_2 & \text{ otherwise}
    \end{array}
  \right.
  , \label{Eq:gamma_inj}
\end{equation}
such that $\gamma_1 \Dw + \gamma_2 (L-\Dw)=0$ with the ALJJ circumference $L=2\pi R$, where $R$ is the mean radius of the ALJJ. In this case, $U(x_0)$ looks like an asymmetric saw-tooth potential with the steep slope $\propto\gamma_1$ and the gentle slope $\propto\gamma_2$, \ie, the asymmetry depends on the width $\Dw$ compared to the junction length $L$, see Fig.~\ref{fig:JJ_Layout_PovRay_bb}. The amplitude of the potential can be varied by changing $I_\mathrm{inj1}=\gamma_1 \Dw w_J$, which, in principle, also allows the operation as a flashing ratchet. Here, we focus on the rocking ratchet only, \ie, the potential $\propto \gamma_\mathrm{inj1}$ is (almost) constant and $\xi(t)\neq 0$. To apply the current $I_\mathrm{inj1}$ we use one of the two injectors visible in Fig.~\ref{fig:JJ_Layout_PovRay_bb}. The injector is attached to the bottom electrode. The current $I_\mathrm{inj1}$ is injected via inj1 into the bottom electrode of the ALJJ and is used to create the asymmetric potential as described above.

The periodicity of the potential is provided by using an annular LJJ (ALJJ) and $\av[x]{\gamma_\mathrm{inj}(x)}=0$. Note, that the latter condition is automatically satisfied because $I_\mathrm{inj1}$ is applied to the same electrode so that $\gamma_1 j_c \Dw w_J = \gamma_2 j_c (L-\Dw) w_J = I_\mathrm{inj1}$.

In addition, our JVR has a pair of current injectors (further called inj2) separated by a distance $\Delta x$ and attached to the top superconducting electrode, see Fig.~\ref{fig:JJ_Layout_PovRay_bb}. They are used to insert a fluxon (Josephson vortex) in the ALJJ \cite{Ustinov02a, Malomed04}.

\section{Samples}
\label{sec:Samples}

We investigated several Nb/Al-AlO$_x$/Nb junctions with different parameters. Here, we report the results obtained for two samples with parameters summarized in Table~\ref{tab:ParametersOfTheUsedJunctions}.
\begin{table}[t]
  \centering
    \begin{tabular}{r r r r r r}
    \hline
    \hline
    Sample & $r$ & $j_c$ & $\lambda_\mathrm J$ & $l$ & $\Delta x=\Delta w$
    \\
    & [$\units{\mu m}$] & [\units{A/cm^2}] & [\units{\mu m}] &  & [\units{\mu m}] \\
    \hline
    C3 & \multicolumn{1}{r}{70} & \multicolumn{1}{r}{87} & \multicolumn{1}{r}{47} & \multicolumn{1}{r}{9.4} & \multicolumn{1}{r}{5} \\
    E3 & \multicolumn{1}{r}{30} & \multicolumn{1}{r}{138} & \multicolumn{1}{r}{29} & \multicolumn{1}{r}{6.5} & \multicolumn{1}{r}{2} \\ \hline
    \hline
    \end{tabular}
  \caption{Parameters of the used junctions. $r$ is the junction radius, $j_c$ is the critical current density at $4.2\,\textrm{K}$, $\lambda_\mathrm J$ is the Josephson penetration depth, $l$ is the normalized junction length. $\Delta x$ and $\Delta w$ describe the injector seperation and width (see text).}
  \label{tab:ParametersOfTheUsedJunctions}
\end{table}
For all samples, $\lambda_J \gg w_J=5\units{\mu m}$ and $\lambda_J \gg \Delta w=\Delta x$ (see Table~\ref{tab:ParametersOfTheUsedJunctions}), \ie, we can treat our ALJJs as one-dimensional and inj2 as an ideal discontinuity \cite{Malomed04}. The maximum revolution frequency for a fluxon inside an ALJJ equals $\nu_0=\overline{c}_0/L$ with the Swihart velocity $\overline{c}_0$. The corresponding voltage (voltage of the first fluxon step) is given as
\begin{equation}
  V_1=\Phi_0\nu_0=\Phi_0\omega_p/l
  \label{eqn:FluxonVoltageWithDrive}
\end{equation}
with the normalized length $l=L/\lambda_J$.

\begin{figure}[t!]
  \centering
    \includegraphics{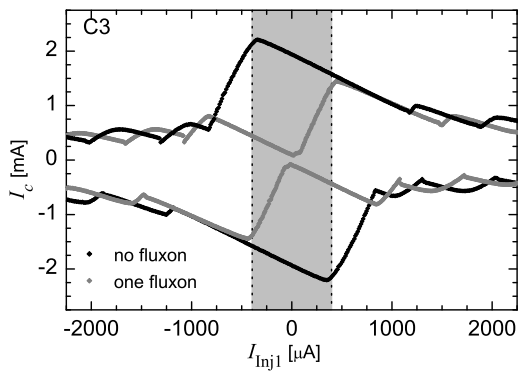}
  \caption{%
    $I_c(I_\mathrm{inj1})$ dependence for sample C3. The range of $I_\mathrm{inj1}$ relevant for ratchet operation, i.e., the region where the picture of a fluxon-particle moving in a potential is valid, is shaded.
  }
  \label{fig:Ic(I_Inj1)_bb}
\end{figure}
All measurements have been performed at the temperature \mbox{$T=4.2\,\mathrm{K}$}. All junctions showed good $I$--$V$ characteristics (IVC) and symmetric $I_c(H)$ dependences (not shown). Their $I_c(I_\mathrm{inj2})$ dependences look Fraunhofer-like in agreement with the theory \cite{Malomed04}. The first minimum of this dependence corresponds to the phase being twisted by $\pm 2\pi$ in a tiny region between the injector pair inj2 and to a free (anti)fluxon being inserted into the ALJJ region outside inj2. Thus, we insert a fluxon into the junction by choosing the corresponding value of $I_\mathrm{inj2}$.

To calibrate inj1, we measure $I_c(I_\mathrm{inj1})$ (see Fig.~\ref{fig:Ic(I_Inj1)_bb}). By measuring $I_c(I_\mathrm{inj1})$ without a fluxon inside the junction ($I_\mathrm{inj2}=0$) and comparing it to $I_c(I_\mathrm{inj1})$ measured with a fluxon inside the junction ($I_\mathrm{inj2}\propto\Delta\phi=\pm 2\pi$), the asymmetry of the ratchet is determined and the range of magnetic field (range of the amplitude of the ratchet potential) relevant to the ratchet operation is found \cite{Goldobin2001, Beck2005}, see Fig.~\ref{fig:Ic(I_Inj1)_bb}. Inside this working range, the depinning current $I_c$ scales almost linearly with $I_\mathrm{inj1}$, but it is asymmetric for positive and negative direction of the bias current $I$ (driving force). Note, that like in Ref.~\cite{Beck2005} a residual pinning of the fluxon due to the finite inj2 sizes $\Delta x$ and $\Delta w$ is still visible in Fig.~\ref{fig:Ic(I_Inj1)_bb} \cite{Malomed04}.

Measuring an IVC with a fluxon inside the junction and $I_\mathrm{inj1}\neq 0$, \ie, an applied ratchet potential, a fluxon step appears on the IVC (not shown) corresponding to the rotation of a fluxon around the ALJJ with $u\approx\overline{c}_0$. The depinning currents $I_c$ and return currents $I_r$ of the fluxon step depend on the polarity of the applied bias as well as on $I_\mathrm{inj1}$.

\section{Quasi-static drive}
\label{sec:QuasiStaticDrive}

\begin{figure}[tb]
  \begin{center}
    \includegraphics{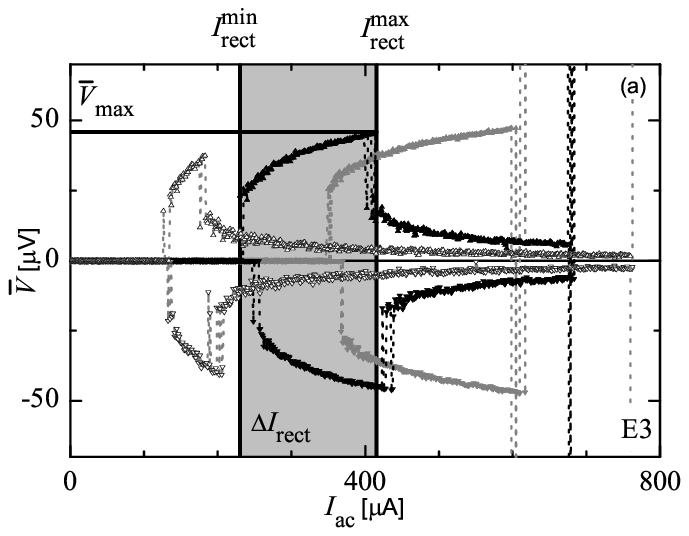}
    \includegraphics{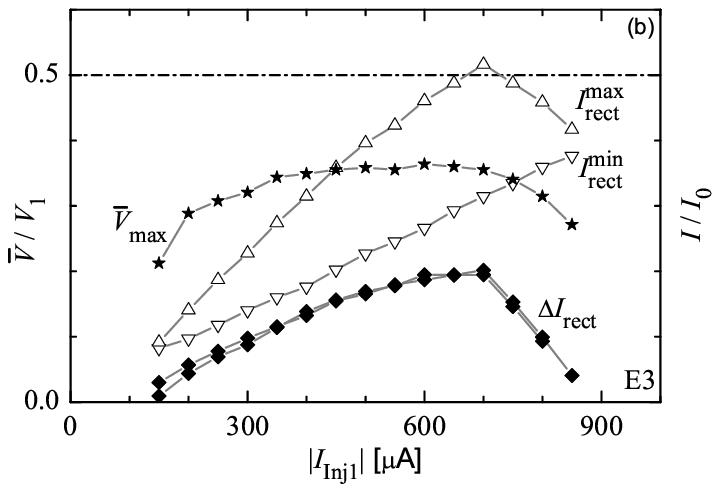}
  \end{center}
  \caption{%
    (a) Typical rectification curves $\mean{V}(\Iac)$ for $I_\mathrm{inj1}=\pm 200\units{\mu A}$, $\pm 400\units{\mu A}$ and $\pm 600\units{\mu A}$ (open, black and grey symbols, respectively).
    (b) Figures of merit for different potential amplitudes normalized to $I_0$ and $V_0$, respectively. $\Delta I_\mathrm{rect}$ is shown for both positive and negative potential amplitude.
  }
  \label{fig:V(I_rf)_qs_bb}
\end{figure}

For our experiments in the quasi-static regime, we apply a periodic bias current \mbox{$I(t)=\xi(t)j_c L w_J = \Iac\sin(2\pi\nu t)$} with the frequency \mbox{$\nu=100\,\mathrm{Hz}\ll\nu_0$} and measure the rectification curve $\mean{V}(\Iac)$ by averaging the voltage over $10\,\mathrm{ms}$ ($1000$ data points sampled at $100\,\mathrm{kHz}$) --- one period of the ac drive. For the junction E3 the dependence $\mean{V}(\Iac)=V_1\cdot\mean{u}(\Iac)$ is shown in Fig.~\ref{fig:V(I_rf)_qs_bb}~(a) for different values of $I_\mathrm{inj1}$, \ie, for different amplitudes of the potential. Note, that the absolute maximum $\mean{V}$, that can be obtained with a rectangular-shaped driving force, is given by $0.5\cdot V_1$, because the fluxon is driven only half a period in one direction. For a sine-like driving force, as in our case, the maximum voltage is somewhat smaller \cite{Goldobin2001}.

All $\mean{V}\left( \Iac\right)$ curves show similar features. For small $\Iac$ the driving force acting on a fluxon is not sufficient to push the fluxon out of the potential well in either direction so that $\mean{u} \propto \mean{V}=0$. At $\Iac>I_\mathrm{rect}^\mathrm{min}$ the bias is able to push the fluxon in one direction but not in the other, which results in $\mean{u} \propto \mean{V}\neq 0$. At $\Iac>I_\mathrm{rect}^\mathrm{max}$, the junction switches into the resistive state generating a high positive or negative dc voltage. The latter regime is not discussed here as it has nothing to do with the JVR operation. Rarely, we also observed the typical ratchet behavior -- a decrease of $\mean{V}$ at $\Iac>I_\mathrm{rect}^\mathrm{max}$ when the driving force is able to overcome the potential barrier in both directions and rectification, thus, drops significantly. We rarely observe this regime since the asymmetry is so large that a negative fluxon step does not appear in most cases.

For not very large potential heights $\propto I_\mathrm{inj1}$, i.e., when the perturbation theory is applicable, the values $I_\mathrm{rect}^\mathrm{min}$ and $I_\mathrm{rect}^\mathrm{max}$ grow approximately linearly with $I_\mathrm{inj1}$, see Fig.~\ref{fig:V(I_rf)_qs_bb}(b). Therefore, the size of the rectification window $\Delta I_\mathrm{rect}=I_\mathrm{rect}^\mathrm{max}-I_\mathrm{rect}^\mathrm{min}$ also grows approximately linearly with $I_\mathrm{inj1}$. Further, $V_\mathrm{max}$ grows with $I_\mathrm{inj1}$, but it reaches its maximum value already for medium values of $I_\mathrm{inj1}$, see Fig.~\ref{fig:V(I_rf)_qs_bb}(b). At high potential amplitudes the rectification window is becoming smaller, probably because the ratchet potential cannot be considered anymore as a perturbation.

The dependence shown in Fig.~\ref{fig:V(I_rf)_qs_bb}(b) is qualitatively the same for all measured samples and for both signs of the potential, as long as the junction length $l$ is large enough. For $l\lesssim 4$ the asymmetry of the potential is strongly reduced because the potential is a convolution of the magnetic field profile created by inj1 and the fluxon shape\cite{Goldobin2001}. The main result of Fig.~\ref{fig:V(I_rf)_qs_bb}(b) is that in order to obtain large $V_\mathrm{max}$ and large $\Delta I_\mathrm{rect}$, one should operate the ratchet at large amplitudes of the potential, \ie, at large values of $I_\mathrm{inj1}$.

\begin{figure}[!tbp]
  \begin{center}
    \includegraphics{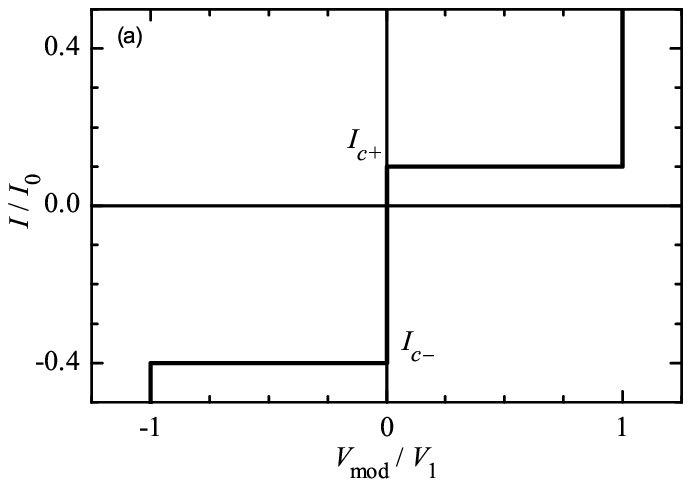}\\[2mm]
    \includegraphics{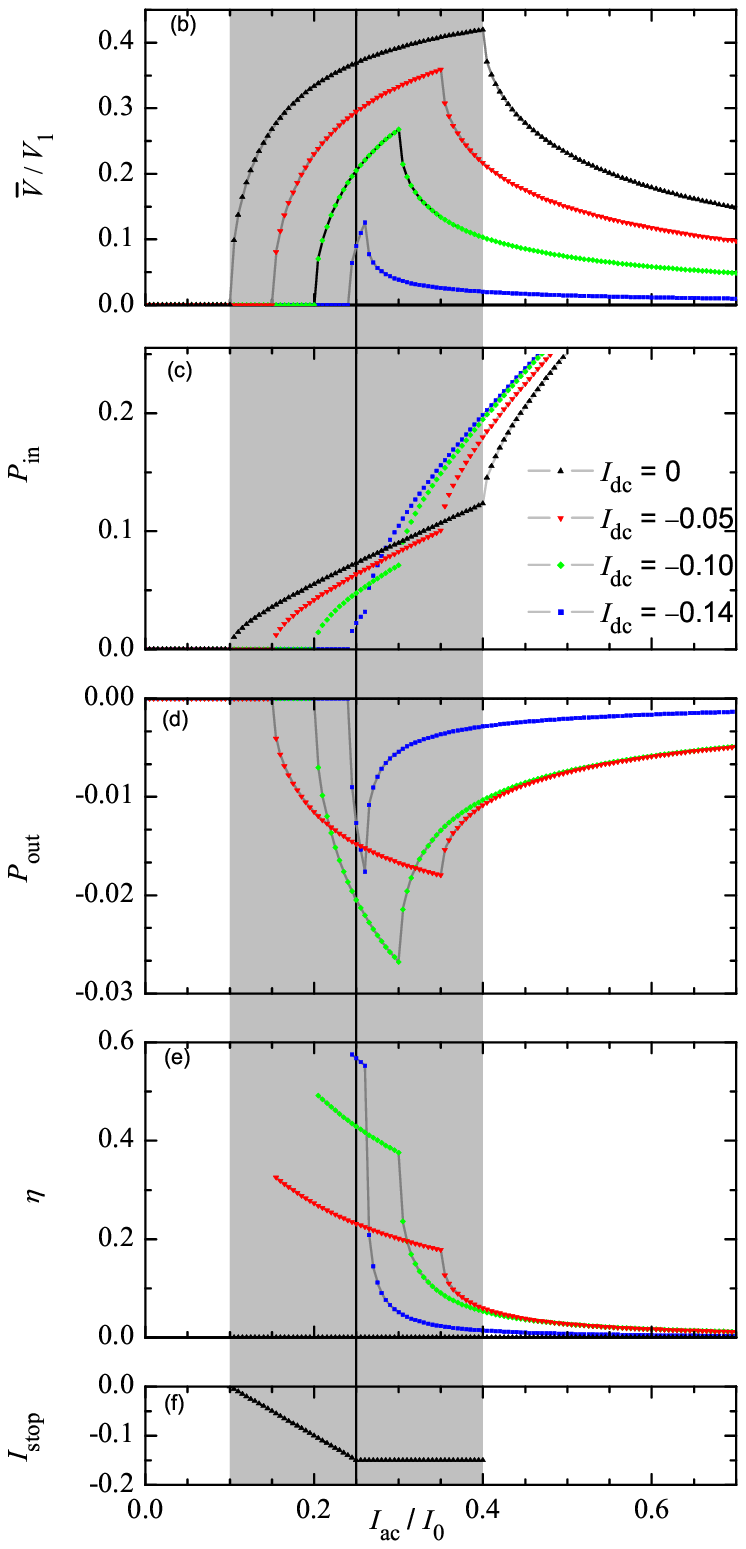}
  \end{center}
  \caption{(Color online)
    (a) Model step-function-like IVC \eqref{Eq:Anal:IVC} with $I_{c+}=0.1 I_0$ and $I_{c-}=-0.4 I_0$,  $I_0=j_c w_J L$ is the intrinsic critical current. (b)  $\mean{V}\left( \Iac\right)$ curves calculated using Eq.~\eqref{Eq:Anal:meanV}. (c) $P_\mathrm{in}(\Iac)$ \eqref{Eq:Anal:Pin}. (d) $P_\mathrm{out}(\Iac)$ \eqref{Eq:Pout}. (e) power efficiency $\eta=-P_\mathrm{out}/P_\mathrm{in}$ for different values of $\Idc=0,\,0.05,\,0.10,\,0.14$.
    (f) $\Istop(\Iac)$, see Eqs.~\eqref{Eq:Anal:I_stop-lin} and \eqref{Eq:Anal:I_stop-sat} with saturation value $\Istop=-0.15$ \eqref{Eq:Anal:I_stop-sat}.
  }
  \label{Fig:AnalModel}
\end{figure}

In the quasi-static regime, all information about the ratchet operation can be derived from its IVC. Therefore, we can derive all figures of merit that we are interested in by using a model step-like IVC shown in Fig.~\ref{Fig:AnalModel}(a), which is very similar to a real one. In this IVC that we express as
\begin{equation}
  V_\mathrm{mod}(I) = V_1 \cdot \left\{
  \begin{array}{rrcl}
  -1  &        & I & < \Icn\\
   0, & \Icn < & I & < \Icp\\
  +1, & \Icp < & I &
  \end{array}\right.
  , \label{Eq:Anal:IVC}
\end{equation}
the fluxon depinning currents $\Icp>0$ and $\Icn<0$ have different values reflecting the asymmetry of the potential. The fluxon step is roughly approximated by a vertical step with infinite height. An applied current $I(t)=\Idc+\Iac\sin\left(\omega t\right)$ consists of a dc current and an ac current with frequency $\omega$.

To obtain a rectification curve $\mean{V}(\Iac)$ we integrate the instant voltage over one period $T=2\pi/\omega$ of ac drive, i.e.,
\begin{equation}
  \mean{V}(\Iac) = \frac{1}{T}\int_{0}^{T} V_\mathrm{mod}[I(t)]\, dt
  .\label{Eq:NumRectCurve}
\end{equation}
Within our simple model, this can be integrated explicitely resulting in
\begin{equation}
  \mean{V} = \left\{
  \begin{array}{lrcl}
  0,          &             &\Iac& < \Icp-\Idc\\
  V_{+},      & \Icp-\Idc < &\Iac& < \Idc-\Icn\\
  V_{+}+V_{-},& \Idc-\Icn < &\Iac&
  \end{array}\right.
  , \label{Eq:Anal:meanV}
\end{equation}
where
\begin{subequations}
  \begin{eqnarray}
    V_{+}&=&\frac{+1}{2\pi}\left[ \pi + 2 \arcsin\left( \frac{\Idc-I_{c+}}{\Iac} \right) \right]
    ; \label{Eq:V-}\\
    V_{-}&=&\frac{-1}{2\pi}\left[ \pi - 2 \arcsin\left( \frac{\Idc-I_{c-}}{\Iac} \right) \right]
    , \label{Eq:V+}
  \end{eqnarray}
\end{subequations}
describe the part of $\mean{V}$ rectified during the positive and the negative semi-period of ac drive, respectively.

The $\mean{V}(\Iac)$ calculated in this way is shown in Fig.~\ref{Fig:AnalModel}(b). First, for $\Idc=0$ our simple model gives $I_\mathrm{rect}^\mathrm{min}=I_{c+}$, $I_\mathrm{rect}^\mathrm{max}=-I_{c-}$ and $\Delta I_\mathrm{rect}=-I_{c-}-I_{c+}$. For the chosen values of $|I_{c-}|>|I_{c+}|$ the rectified voltage $\mean{V} \geq 0$.

Second, we apply $\Idc$ to try to stop the ratchet operation at a given value of $\Iac$. The sign of $\Idc$ should be opposite to the sign of $\mean{V}$, i.e. $\Idc< 0$ in our case. Rectification curves $\mean{V}(\Iac)$ calculated using Eq.~\eqref{Eq:Anal:meanV} at different values of $\Idc<0$ are also shown in Fig.~\ref{Fig:AnalModel}(b). One can see that the rectification window shrinks, i.e., close to the edges of the original window the ratchet is not strong enough to work against $\Idc$. With applied $\Idc$
\begin{eqnarray}
  I_\mathrm{rect}^\mathrm{min}&=&\Icp-\Idc
  ; \label{Eq:I_rect^min}\\
  I_\mathrm{rect}^\mathrm{max}&=&\Idc-\Icn
  ; \label{Eq:I_rect^max}\\
  \Delta I_\mathrm{rect}&=&2\Idc-\Icn-\Icp
  , \label{Eq:RectWin}
\end{eqnarray}
which can also be seen from Eq.~\eqref{Eq:Anal:meanV}.

One can also take a different point of view and study how strong the ratchet is at each particular \Iac, i.e., one varies $\Idc$ at fixed $\Iac$. The value of $\Idc$ at which the ratchet stops moving or starts moving backwards ($\mean{V}=0$ or changes sign) is called stopping force (or stopping current) $\Istop$. The dependence  $\Istop\left(\Iac\right)$ is shown in Fig.~\ref{Fig:AnalModel}(f). One can see, that naturally $\Istop=0$ up to the driving amplitude $\Iac=I_{c+}$, because the ratchet didn't even start working yet, so no force is needed to stop it. Then $\Istop$ grows linearly to a value of $\Istop^\mathrm{max}=\Delta I_\mathrm{rect}/2=(-I_{c-}-I_{c+})/2$, which is reached at $\Iac=(-I_{c-}+I_{c+})/2$, i.e., at the middle of the rectification window. In this regime, the ratchet rectifies because during some fraction of the positive semi-period $I(\tau)$ exceeds $I_{c+}$ and never exceeds $I_{c-}$, see Fig.~\ref{Fig:AnalModel}(a).
An additional $\Idc$ ``shifts'' the origin of ac oscillations down along the $I$-axis. Thus, for $\Idc<|\Iac|-\Icp$, the ac sweep does not exceed the depinning current $\Icp$ and rectification vanishes. Therefore
\begin{equation}
  \Istop = |\Iac|-\Icp
  \text{ for } |\Iac|>\Icp
  , \label{Eq:Anal:I_stop-lin}
\end{equation}
which corresponds to the linear part in Fig.~\ref{Fig:AnalModel}(f). The behavior described by Eq.~\eqref{Eq:Anal:I_stop-lin} does not hold for arbitrarily large \Iac. When $|\Iac|>(I_{c+}-I_{c-})/2$, at some amplitude of $\Idc$ the ac sweep also exceeds the negative depinning current $I_{c-}$, which results in a drastic decrease of rectification and also $I_\mathrm{rect}^\mathrm{max}$ given by Eq.~\eqref{Eq:I_rect^max}, see Fig.~\ref{Fig:AnalModel}(b). At $\Idc=(I_{c-}+I_{c+})/2$, the ac sweep sees symmetric depinning currents and rectification vanishes. Thus
\begin{equation}
  \Istop = (I_{c-}+I_{c+})/2 \text{ for }|\Iac|>(I_{c+}-I_{c-})/2
  , \label{Eq:Anal:I_stop-sat}
\end{equation}
corresponding to a saturation of $\Istop$ in Fig.~\ref{Fig:AnalModel}(f) and a completely closed
rectification window on the $\mean{V}(\Iac,\Istop)$ plot. For $\Idc<(I_{c-}+I_{c+})/2$, the rectified voltage $\mean{V}$ changes sign.

Let us now discuss a power balance in our ratchet. Since $\Idc=const.$, the average dc output power
\begin{equation}
  P_\mathrm{out}=\frac{1}{T}\int_0^T V(t) \Idc\, dt
  = \mean{V} \Idc \leq 0
  , \label{Eq:Pout}
\end{equation}
where $\mean{V}(\Iac,\Idc)$ is given by Eq.~\eqref{Eq:Anal:meanV}. $P_\mathrm{out}$ is negative because the ratchet delivers the power to the dc source (load) instead of consuming it. Simultaneously, the input power is given by
\begin{equation}
  P_\mathrm{in} = \frac{1}{T}\int_0^T V(t) \Iac\sin(\omega t)\, dt \geq 0
  . \label{Eq:Pin}
\end{equation}
Within our simple model, this can be integrated explicitely resulting in
\begin{equation}
  P_\mathrm{in} = \left\{
  \begin{array}{lrcl}
  0,                                  &             &\Iac& < \Icp-\Idc\\
  P_\mathrm{in}^{+},                  & \Icp-\Idc < &\Iac& < \Idc-\Icn\\
  P_\mathrm{in}^{+}+P_\mathrm{in}^{-},& \Idc-\Icn < &\Iac&
  \end{array}\right.
  , \label{Eq:Anal:Pin}
\end{equation}
where
\begin{subequations}
  \begin{eqnarray}
    P_\mathrm{in}^{+}=\frac{1}{\pi \Iac} \sqrt{\Iac^2-(\Idc-\Icp)^2}
    ; \label{Eq:P-}\\
    P_\mathrm{in}^{-}=\frac{1}{\pi \Iac} \sqrt{\Iac^2-(\Idc-\Icn)^2}
    , \label{Eq:P+}
  \end{eqnarray}
\end{subequations}
describe the part of $P_\mathrm{in}$ consumed during the positive and the negative semi-period of ac drive, respectively.

Both $P_\mathrm{in}(\Iac)$ and $P_\mathrm{out}(\Iac)$ at different values of $\Idc$ are shown in Fig.~\ref{Fig:AnalModel}(c) and (d). One can see that $P_\mathrm{in}(\Iac)$ has two characteristic branches corresponding to the dissipation during the positive semi-period and during both the positive and negative semi-periods. In contrast, $P_\mathrm{out}(\Iac)$ is non monotonous and has an extremum given by
\begin{equation}
  \mean{V}\left[ I_\mathrm{rect}^\mathrm{max}(\Idc) \right] \Idc
  = \Idc \left[
    \frac12 + \frac1\pi \arcsin\left( \frac{\Idc-I_{c+}}{\Idc-I_{c-}} \right)
  \right]
    \label{Eq:Vmax(Idc)}
\end{equation}
at $\Iac=I_\mathrm{rect}^\mathrm{max}(\Idc)$ \eqref{Eq:I_rect^max}, which depends on $\Idc$ nonmonotonously. One can see that $P_\mathrm{out}=0$ not only at $\Idc=0$, but also at $\Idc=\Istop$ where $\mean{V}=0$, c.f. Eq.~\eqref{Eq:Pout}. Therefore, the maximum power is reached for some intermediate values of $\Istop<\Idc<0$. The exact value can be derived by looking for the extremum of expression \eqref{Eq:Vmax(Idc)} with respect to $\Idc$. It is reached for $\Idc^\mathrm{opt}$ which is a solution of the following transcendental equation
\begin{widetext}
\begin{equation}
  \left( \frac{I_{c+}-I_{c-}}{\Idc^\mathrm{opt}-I_{c-}} \right)
  \frac{\Idc^\mathrm{opt}}{\pi}
  + \left[  \frac12 +
    \frac1\pi\arcsin\left( \frac{\Idc^\mathrm{opt}-I_{c+}}{\Idc^\mathrm{opt}-I_{c-}} \right)
  \right]
  \sqrt{1-
    \left(
      \frac{\Idc^\mathrm{opt}-I_{c+}}{\Idc^\mathrm{opt}-I_{c-}}
    \right)^2
  } = 0
  . \label{Eq:Pmax}
\end{equation}
\end{widetext}
For our parameters, $\Idc^\mathrm{opt}\approx -0.104 I_0$ and $P_\mathrm{max}(\Idc^\mathrm{opt})=-0.028 V_1 I_0$.

Another important figure of merit is the efficiency, defined as  $\eta=-P_\mathrm{out}/P_\mathrm{in}$. The plots $\eta(\Iac)$ are shown in Fig.~\ref{Fig:AnalModel}(e). Obviously, the efficiency has a maximum just at the beginning of the rectification window and falls with increasing $|\Idc|$. To derive the approximate behavior of $\eta(\Iac)$ at the beginning of rectification window analytically, we Taylor-expand $P_\mathrm{out}$ and $P_\mathrm{in}$ near $\Iac=I_\mathrm{rect}^\mathrm{min}(\Idc)$ \eqref{Eq:I_rect^min}. As a result, we get
\begin{equation}
  \eta(\Iac,\Idc)
  \approx \frac{\Idc}{\Idc-I_{c+}}
  - \frac1{6}\frac{\Idc}{(\Idc-I_{c+})^2}(\Iac-I_\mathrm{rect}^\mathrm{min} )
  . \label{Eq:eta(Iac,Idc)}
\end{equation}
The first term represents the exact expression for the efficiency at the left edge of the rectification window for given $\Idc$. To find the maximum efficiency, we vary $\Idc$ from 0 down to $\Istop^\mathrm{max}=(I_{c+}+I_{c-})/2$. At $\Idc \to \Istop^\mathrm{max}$, the efficiency approaches it ultimate maximum value
\begin{equation}
  \eta_\mathrm{max} = \frac{I_{c-}+I_{c+}}{I_{c-}-I_{c+}}
  , \label{Eq:Eff_max}
\end{equation}
although the rectification window $\Delta I_\mathrm{rect}$ vanishes, see Fig.~\ref{Fig:AnalModel}(b)--(e). For our values of $I_{c-}$ and $I_{c+}$ (see the caption of Fig.~\ref{Fig:AnalModel}) $\eta_\mathrm{max}=0.6$.

\section{Non-adiabatic drive}
\label{sec:NonAdiabaticDrive}
We use a microwave generator with an emitting antenna close to the ALJJ to drive the ratchet at frequencies as high as $\nu\gtrsim 1\,\mathrm{GHz}$. With this geometry, the bias leads of the ALJJ act as a pickup antenna. Note that the microwaves are also picked up by the injector electrodes but due to the geometry of the setup it causes only a negligible ``flashing'' part in our ratchet behavior \cite{Beck2005}. We average the dc voltage $\mean{V}$ over $2000$ data points at a sampling rate of $100\,\mathrm{kHz}$ ($\sim 10^8$\,periods) and measure it vs. the applied power $P\propto \Iac$ of the generator.

Rather than growing smoothly as in Fig.~\ref{fig:V(I_rf)_qs_bb}(a), the rectified voltage \mean{V} is now quantized as
\begin{equation}
  \mean{V}_n=-\sgn{\left(I_\mathrm{inj1}\right)}n\Phi_0\nu
  \label{Eq:Vn}
\end{equation}
(Shapiro-like steps) \cite{Goldobin2001}. Each step corresponds to an integer number $n$ of turns of a fluxon around the ALJJ per one period of the ac drive. The pre-factor $-\sgn{\left(I_\mathrm{inj1}\right)}$ is chosen so that $n>0$ corresponds to the motion of a fluxon in the ``easy'' direction of the potential, while $n<0$ corresponds to the difficult direction (current/voltage reversal), \cf, Fig.~\ref{fig:Ic(I_Inj1)_bb}. If the time $\tau_0$ required for one revolution of a fluxon around the ALJJ (in the best case $\tau_0=\overline{c}_0/L$) becomes comparable with the period of the driving force, \ie, if the voltage $\mean{V}=\Phi_0\nu$ is approaching $V_1$, the fluxon has time for only one revolution, so $n=1$ and only one step can be observed on the $\mean{V}(P)$ curves. In the non-adiabatic regime, also current reversal can be observed, \ie, $n=-1$ resulting in a ``negative'' voltage \cite{Chauwin1995:CR:AssPump,Jung1996,Mateos2000,Mateos2003:BattleOfAttractors,Son2004:UdmpDetRat:CurrRev}.

Being in the rectification regime, we apply an additional bias current $\Idc$ which either tries to stop the ratchet operation (\ie, with sign opposite to $\mean{V}$) or to ``help'' it (\ie, with the same sign as $\mean{V}$). In the non-adiabatic regime it makes sense to associate $\Istop$ and $\Isupp$ with a particular mode of operation (step $n$) rather than with the ratchet as a whole. Therefore, a particular mode of operation is ending when the voltage $\mean{V}\left( \Idc\neq 0\right)\neq \mean{V}\left( \Idc=0\right)$. The value of $\Idc$ at which this condition is met is called \emph{stopping force} $\Istop\left( P\right)$ or \emph{support force} $I_\mathrm{supp}\left( P\right)$ respectively.

\begin{figure}[htb]
  \begin{center}
    \includegraphics{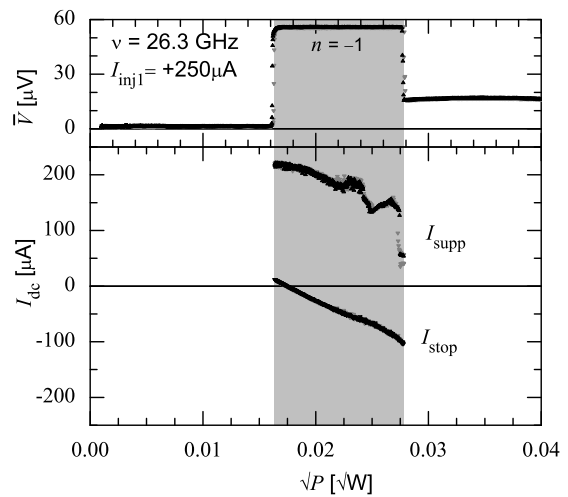}
  \end{center}
  \caption{%
    Rectification curve with positive potential sign. Stopping force $\Istop(P)$ and support force $I_\mathrm{supp}(P)$. Black triangles: $P$ is swept up; grey triangles: $P$ is swept ``down''.
  }
  \label{fig:Istop_26.3GHz_posPot_bb}
\end{figure}

\begin{figure}[htb]
  \begin{center}
    \includegraphics{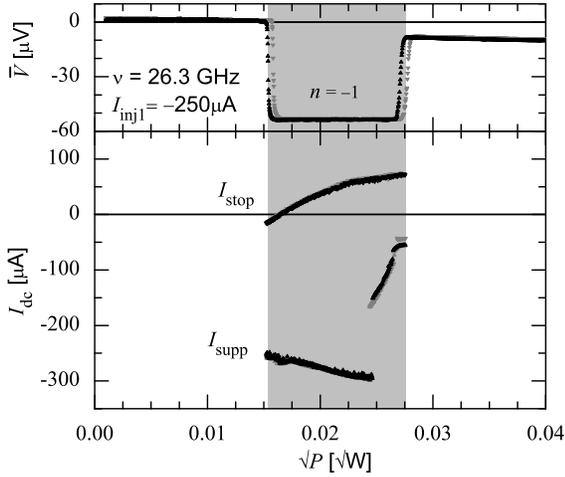}
  \end{center}
  \caption{%
    Rectification curve with negative potential sign. Stopping force $\Istop\left( P\right)$ and support force $I_\mathrm{supp}(P)$. Black triangles correspond to sweep up, grey triangles to sweep down.
  }
  \label{fig:Istop_26.3GHz_negPot_bb}
\end{figure}

Fig.~\ref{fig:Istop_26.3GHz_posPot_bb} shows a typical rectification curve measured using the sample C3 at $\Idc=0$. The voltage step corresponds to $n=-1$.
%
In addition, we show the $\Istop(P)$ and $I_\mathrm{supp}(P)$ dependences for this area. Fig.~\ref{fig:Istop_26.3GHz_negPot_bb} shows the curves for the same parameter setting, only the potential amplitude is being reversed in comparison to Fig.~\ref{fig:Istop_26.3GHz_posPot_bb}.

The rectification curves $\mean{V}(\Iac)$ show the expected discrete values of rectified voltage. An inverted potential causes an inverted voltage step. Different sweeping directions of $P$ result in almost identical curves, which suggests that we are not observing a simple synchronization of the fluxon movement with the drive (for which we can lock to the $n=\pm1$ step randomly) but rather a true rectification with well defined direction independent from the history. We check this for every voltage step to ensure that we are observing a real ratchet effect and also measure $\Istop$ and $I_\mathrm{supp}$ for different sweeping directions to avoid regimes of synchronization.

In both plots, $\Istop=0$ at the lower edge of the rectification window --- a feature that we observe in almost all measurements. From this point, the stopping force $\Istop$ grows smoothly throughout the rectification window. It vanishes when we leave the rectification regime of $n=-1$.

The support force $I_\mathrm{supp}$ also looks similar in both plots (but with opposite sign). It has a minimum value at the upper edge of rectification window. At lower driving amplitudes $\Isupp$ stays $\approx const$. and drops very fast (or even jumps abruptly) down to its minimum at the upper edge of the rectification window.

The maximum value of $\Istop$ in all our measurements was $|\Istop^\mathrm{max}|=215\units{\mu A}=0.112 I_0$ (reached at the edge of the rectification window) and $|I_\mathrm{supp}^\mathrm{max}|=298\units{\mu A}=0.155 I_0$ (both in sample C3, not shown). Note, that $|I_\mathrm{supp}^\mathrm{max}|>|\Istop^\mathrm{max}|$ in all measurements.

\begin{figure}[t]
  \begin{center}
    \includegraphics{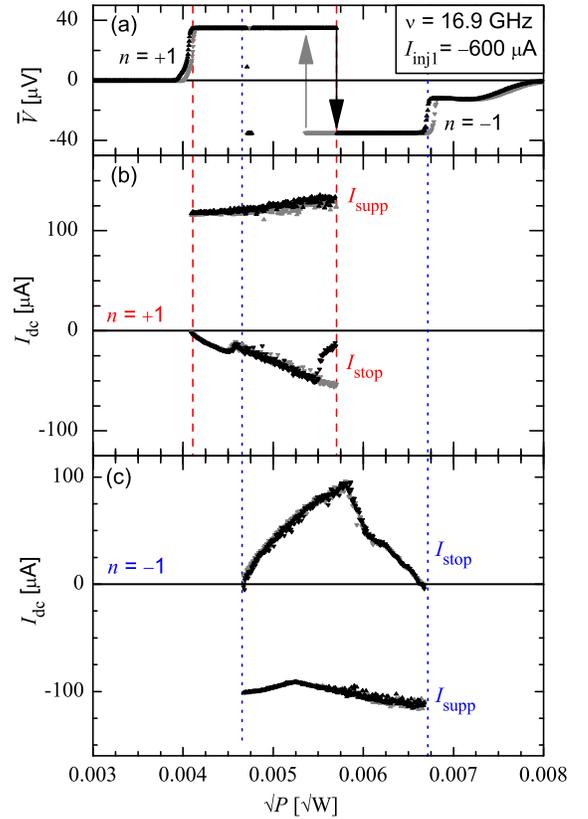}
  \end{center}
  \caption{(Color online)
    Rectification curve for $\nu=16.9\,\mathrm{GHz}$ (a). Stopping force $\Istop\left( P\right)$ and support force $I_\mathrm{supp}\left( P\right)$ for voltage step corresponding to $n=+1$ (b) and $n=-1$ (c).
  }
  \label{fig:Istop_16.9GHz_bb}
\end{figure}

In sample E3, we observed both direct rectification ($n=+1$) and reversal ($n=-1$) in one rectification curve at a driving frequency $\nu=16.9\,\mathrm{GHz}$, see Fig.~\ref{fig:Istop_16.9GHz_bb}(a). The voltage jumps directly from one step to the other, showing a small hysteresis (indicated by the arrows) when sweeping $P$ back and forth. We measured $\Istop(P)$ and $I_\mathrm{supp}(P)$ for both voltage steps. Also note a data point at $\sqrt{P}\approx0.0048\,\sqrt{\textrm{W}}$. The step with $n=-1$ is metastable in the interval $\sqrt{P}=0.0054\ldots0.0057\,\sqrt{\textrm{W}}$, so one can see the system jump on it sometimes.

Fig.~\ref{fig:Istop_16.9GHz_bb}(b) shows $\Istop(P)$ and $\Isupp(P)$ for $n=+1$. The shape of $\Istop(P)$ is similar to the previous plots. In the middle of the rectification window, the curve gets more noisy and, at the upper edge, $\Istop$ shows a small hysteresis for different sweep directions of $P$. $I_\mathrm{supp}(P)$ is almost constant throughout the major part of the plot and jumps close to the upper edge of the rectification window.

Fig.~\ref{fig:Istop_16.9GHz_bb}(c) shows $\Istop$ and $\Isupp$ for the step $n=-1$. $I_\mathrm{supp}(P)$ has a well-known shape --- the curve remains at high values within the whole rectification window.
The $\Istop\left( P\right)$ curve vanishes at \emph{both} ends of the rectification window. In fact, one can see that it consists of two branches that join at a value of driving amplitude, where the voltage step $n=-1$ is ending -- \ie, the left branch of $\Istop(P)$ shows rectification with $n=-1$, although the $\mean{V}\left( P\right)$ curve has switched already to $n=+1$ dynamics in this region. The reason is that we are tracking the $n=-1$ mode, which is not visible for $\Idc=0$, but appears for $\Idc\neq0$. The fact that we can track it shows, that the presence of the $n=-1$ mode depends on the history of the system and the value of $\Idc$ in a non-trivial way.

The reason for this might be the measurement itself: For the measurement of $\mean{V}\left( P\right)$, the current $\Idc=\mathrm{const.}\approx 0$ (a small current offset is always possible) whilst for the measurement of $\Istop\left( P\right)$ or $I_\mathrm{supp}\left( P\right)$ the current $\Idc$ is ramped up for each amplitude of the drive $P$ and then set to $\Idc\approx 0$ again for the next value of $P$. This subtle difference can influence the fluxon dynamics inside the junction and therefore the $n=-1$ dynamics can be pertained longer than what was seen in the rectification curve. In this sense, the measurement of the stopping force or support force can be used for further investigation of possible modes of operation while measuring the loading capability of the ratchet.

For the non-adiabatic regime, numerical simulations show a behavior similar to the experiment (for the quasi-static regime, these simulations are very time-consuming and can be avoided by using an analytical approach similar to the one discussed in Sec.~\ref{sec:QuasiStaticDrive}). These simulations were performed using an explicit numerical scheme for Eq.~\eqref{eqn:sine-gordon-equation} with damping coefficient $\alpha=0.1$ (weakly underdamped limit). The numerical technique and simulation software are discussed in detail in \cite{Goldobin2003}.

\begin{figure}[htb]
  \begin{center}
    \includegraphics{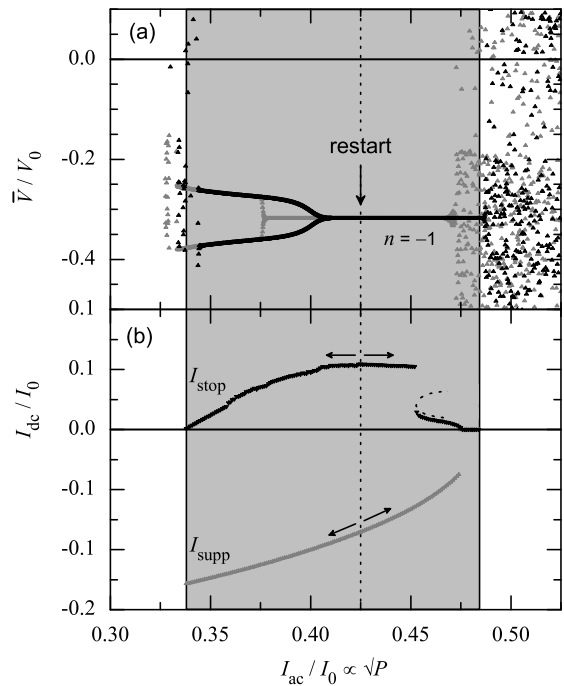}
  \end{center}
  \caption{%
    (a) Numerically simulated $\mean{V}(\Iac)$ for ALJJ of $l=8.0$ and driving frequency $\omega /\omega_\mathrm{pl}=0.25$. Black data points correspond to sweeping \Iac upwards, while the gray points to a downward sweep. The value of \Iac where the simulation is restarted (see text) is indicated by the dotted line.
    (b) $\Istop(\Iac)$ and $I_\mathrm{supp}(\Iac)$ for the voltage step corresponding to $n=-1$.
    Black (gray) symbols correspond to positive (negative) sweeping direction of $\Iac$.
    \Istop is reduced by a region of instability (marked by the dashed line).
  }
  \label{fig:Simu_Istop_l=8_normalRegion_bb}
\end{figure}

In Fig.~\ref{fig:Simu_Istop_l=8_normalRegion_bb} we show simulated $\av{V}(\Iac)$ as well as $\Istop(\Iac)$ and $\Isupp(\Iac)$ dependences. In our simulation of $\mean{V}(\Iac)$ dependences shown in Fig.~\ref{fig:Simu_Istop_l=8_normalRegion_bb}, the sweep of $\Iac$ was performed from $0.3I_0$ to $0.6I_0$ and back. The value of \mean{V} was calculated by averaging $V$ only over \emph{one period} of ac drive. Therefore, if we are in the chaotic regime, we will observe a finite value of the voltage in our simulation, while in the experiment these will be averaged to some finite value which is often equal to zero. The chaotic voltage distribution in Fig.~\ref{fig:Simu_Istop_l=8_normalRegion_bb} at $\Iac>0.48$ shows such a chaotic regime. For this simulation, the (discrete) step in \Iac was chosen very tiny so that the system still has time to come in equilibrium, although one sees the transient processes on $\mean{V}(\Iac)$. 
A large discrete change of the system parameters (\eg the driving power or \Idc) may give a different result than a small one, \ie the resulting value of $\Istop$ or $\Isupp$ can be much smaller than for smooth parameter changes. This happens, because the dynamics is close to chaotic and thus even a small kick (step-like change of parameters) may drive the system into a different state. Our simulations use a small step size of $10^{-5}I_0$.

Then, for the simulation of $\Istop(\Iac)$ and $\Isupp(\Iac)$ we started by recalling the state of the system at $\Iac/I_0=0.425$ (where rectification takes place) and then sweeping $\Iac$ into each direction and, for given \Iac, looking for \Istop or \Isupp. For the interpretation of the plots, the curves at $\Iac<0.425$ have to be interpreted looking at the $\mean{V}(\Iac)$ curve swept in negative direction, while $\Istop(\Iac)$ and $\Isupp(\Iac)$ at $\Iac>0.425$ have to be followed looking at the positive sweep of $\mean{V}(\Iac)$.

The $\mean{V}(\Iac)$-curve reproduces the $n=-1$ step similar to the experiment, \cf Fig.~\ref{fig:Istop_26.3GHz_posPot_bb} and Fig.~\ref{fig:Istop_26.3GHz_negPot_bb}. At $0.34<\Iac<0.41$ one can see period two dynamics in the $\mean{V}(\Iac)$ curve. Again, this can only be observed in simulations as these values are averaged to their mean in our experiment. One can see that the $\Istop(\Iac)$-curve is similarly shaped as in our measurement (see Fig.~\ref{fig:Istop_26.3GHz_negPot_bb}), showing an increasing value to larger driving power. At high driving amplitudes, one can see a jump in the $\Istop(\Iac)$ curve to lower values of \Istop which we did not see in our particular measurement. This jump is related to the presence of an island in the \Idc-\Iac plane (marked by the dashed line) where the dynamics is not $n=-1$ any more. The $\Isupp(\Iac)$-curve shows large values over the whole rectification window. We believe that in our experiment, we also meet the island where the solution for $n=-1$ is not stable.

\begin{figure}[htb]
  \begin{center}
    \includegraphics{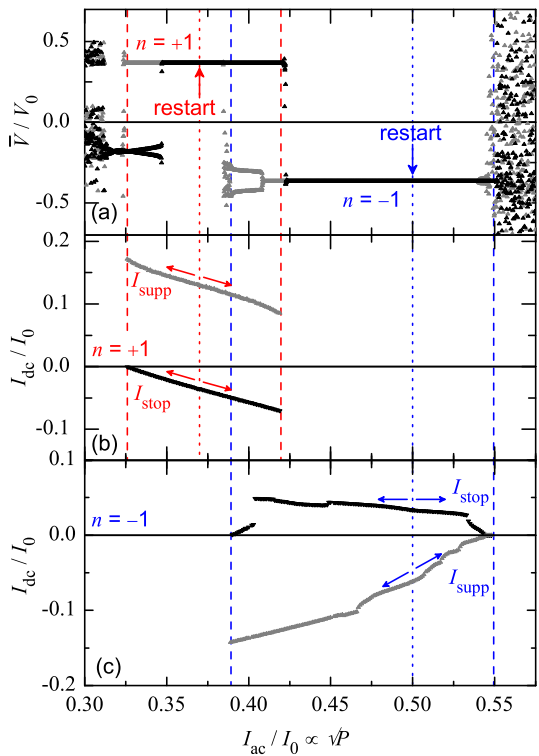}
  \end{center}
  \caption{(Color online)
    (a) Numerically simulated $\mean{V}(\Iac)$ for ALJJ of $l=10.0$ and driving frequency $\omega /\omega_\mathrm{pl}=0.23$. Black data points correspond to sweeping \Iac upwards, while the gray points to a downward sweep.
    (b) $\Istop(\Iac)$ and $I_\mathrm{supp}(\Iac)$ for the voltage step corresponding to $n=+1$.
    Black (gray) symbols correspond to positive (negative) sweeping direction of $\Iac$.
  }
  \label{fig:Simu_Istop_bb}
\end{figure}

We have also been able to reproduce a situation in which both $n=+1$ and $n=-1$ dynamics are present on one rectification curve like in our measurements on sample E3. The corresponding numerical simulation results for a LJJ of the normalized length $l=10.0$ are presented in Fig.~\ref{fig:Simu_Istop_bb}(a).

In this simulation, the sweep of $I_\textrm{ac}$ was performed also from $0.3I_0$ to $0.6I_0$ and back with the same step in $I_\textrm{ac}$ as in Fig.~\ref{fig:Simu_Istop_l=8_normalRegion_bb}(a). For the interpretation of the plots, for $I_\textrm{ac}<0.50I_0$ one has to refer to the $\mean{V}(\Iac)$ curve swept in negative direction, for $\Iac>0.50$ the positive sweep of $\mean{V}(\Iac)$ has to be followed.

In Fig.~\ref{fig:Simu_Istop_bb}(a) a transition between the voltage steps $n=+1$ and $n=-1$ show a small hysteresis.  Moreover, for the negative sweeping direction of $\Iac$, a bifurcation to ``period two'' dynamics is visible. In our experiment with much longer integration time, this is averaged to the same step voltage and is not distinguishable from ``period one'' dynamics. The same applies to the region (outside the $n=\pm1$ voltage steps) where the system shows chaotic dynamics --- due to the averaging in our experiments, one observes just an average voltage of the chaotic fluxon motion.

Fig.~\ref{fig:Simu_Istop_bb}(b) shows the $\Istop(\sqrt{P})$ and $\Isupp(\sqrt{P})$ dependences for $n=+1$ step. They confirm the behavior observed in our measurements. Namely, the stopping force $\Istop(\Iac)$ grows almost linearly from the lower edge of the rectification window to larger values and then jumps down to zero. The support force $I_\mathrm{supp}(\Iac)$ also shows the well-known behavior and drops smoothly to smaller values while approaching the upper edge of the rectification window.

Fig.~\ref{fig:Simu_Istop_bb}(c) shows the $\Istop(P)$ and $\Isupp(P)$ dependences for $n=-1$. They  look very similar to those measured using the sample E3, c.f. Fig.~\ref{fig:Istop_16.9GHz_bb}(c). The stopping force $\Istop\left( \Iac\right)$ also grows from the lower edge of the rectification window and at some point jumps to a branch growing from the upper edge -- in our simulation, the curves also show a jump. This jump is located at a driving amplitude $\Iac$ where the rectification curve $\mean{V}\left( \Iac\right)$ has a bifurcation and is right inside the hysteretic area. The support force $I_\mathrm{supp}\left( \Iac\right)$ again drops to lower values with increasing driving amplitude $\Iac$ and goes down to zero at the upper edge of the rectification window. One can see that the curve consists of several pieces that may be related to the switching onto different dynamic regimes. This is in contrast to the experimental results where $I_\mathrm{supp}\left( \Iac\right)$ is falling abruptly down to zero.

The maximum values $\Istop^\mathrm{max}=0.049I_0$ and $I_\mathrm{supp}^\mathrm{max}=0.167I_0$ in Fig.~\ref{fig:Simu_Istop_bb}(b) and (c) are similar to those obtained in our measurements, other simulations also confirm the scale of these values.

Both in simulation and experiment, the maximum power delivered to the load in the non-adiabatic regime can be obtained by using large driving powers $P$. As shown in Fig.~\ref{fig:Istop_16.9GHz_bb}(c), sometimes the maximum value is reached around the center of the rectification window. Combining these findings, a good value to start is a large value of $P$ on the rectification step with some variations for optimization in the actual experiment. Estimated maximum $P_\mathrm{out}=I_\mathrm{stop}\mean{V}\approx 50\units{\mu A}\times34\units{\mu V}=1.7\units{nW}$ for $n=+1$ and $P_\mathrm{out}\approx 3.3\units{nW}$ for $n=-1$ in Fig.~\ref{fig:Istop_16.9GHz_bb} and $P_\mathrm{out}\approx 5.7\units{nW}$ for $n=-1$ in Fig.~\ref{fig:Istop_26.3GHz_negPot_bb}. Note the difference from the adiabatic case where $\mean{V}$ tends to zero smoothly as a function of \Iac, so that $P_\mathrm{out}\to 0$ too. In the non-adiabatic case the maximum $P_\mathrm{out}$ may be reached at the edge of the step as in Fig.~\ref{fig:Istop_26.3GHz_negPot_bb}.

\section{Conclusion}
\label{sec:Conclusion}

We implemented the Josephson vortex ratchet using an annular long Josephson junction equipped by current injectors used to create the ratchet potential and inject a fluxon into the junction.

In the quasi-static regime we derived the dependence of the stopping force $\Istop\left( \Iac\right)$ analytically using a reasonable model for the IVC. This model predicts that in order to increase the value of the stopping force (and other figures of merit), the amplitude of the ratchet potential should be rather large. The stopping force is directly related to the loading capability of a ratchet and to the maximum rectified power that it can deliver to the load. The maximum output power $P_\textrm{out}$ is also reached at the upper end of the rectification window for intermediate values of $\Idc$. However, the ratchet works with maximum efficiency at the lower end of the rectification window. The ultimate efficiency that can be reached in principle is defined only by the asymmetry of the potential, see Eq.~\eqref{Eq:Eff_max}.

In the non-adiabatic regime, we have measured rectification curves and sometimes observed a transition between $n=+1$ and $n=-1$ dynamics with a small hysteresis. By applying an additional dc bias current $\Idc$ in the rectification regime we have measured the stopping force. Both $\Istop$ and $P_\mathrm{out}$ have maximum near the upper edge of the rectification window. The support force, \ie, the current with the same sign compared to the rectification voltage, where the ratchet stops working, is not changing very drastically throughout the rectification window. 

Therefore, if one wants to have a robust ratchet operation tolerant to additional bias currents, the ratchet should be operated at the upper end of the rectification window and the dc bias current (flowing to the load) should not exceed $\Istop\left( \Iac\right)$.

\acknowledgments
This work was supported by the Deutsche Forschungsgemeinschaft (KO1303/7-1). M. Knufinke acknowledges support by the Carl Zeiss Stiftung.

\bibliography{MyBib}

\begin{thebibliography}{31}
\expandafter\ifx\csname natexlab\endcsname\relax\def\natexlab#1{#1}\fi
\expandafter\ifx\csname bibnamefont\endcsname\relax
  \def\bibnamefont#1{#1}\fi
\expandafter\ifx\csname bibfnamefont\endcsname\relax
  \def\bibfnamefont#1{#1}\fi
\expandafter\ifx\csname citenamefont\endcsname\relax
  \def\citenamefont#1{#1}\fi
\expandafter\ifx\csname url\endcsname\relax
  \def\url#1{\texttt{#1}}\fi
\expandafter\ifx\csname urlprefix\endcsname\relax\def\urlprefix{URL }\fi
\providecommand{\bibinfo}[2]{#2}
\providecommand{\eprint}[2][]{\url{#2}}

\bibitem[{\citenamefont{Feynman et~al.}(1966)\citenamefont{Feynman, Leighton,
  and Sands}}]{Feynman1966}
\bibinfo{author}{\bibfnamefont{R.~P.} \bibnamefont{Feynman}},
  \bibinfo{author}{\bibfnamefont{R.~B.} \bibnamefont{Leighton}},
  \bibnamefont{and} \bibinfo{author}{\bibfnamefont{M.}~\bibnamefont{Sands}},
  \emph{\bibinfo{title}{The Feynman Lectures On Physics}},
  vol.~\bibinfo{volume}{1} (\bibinfo{publisher}{Addison-Wesley Publishing
  Company}, \bibinfo{year}{1966}).

\bibitem[{\citenamefont{Magnasco}(1993)}]{Magnasco1993}
\bibinfo{author}{\bibfnamefont{M.~O.} \bibnamefont{Magnasco}},
  \bibinfo{journal}{Phys. Rev. Lett.} \textbf{\bibinfo{volume}{71}},
  \bibinfo{pages}{1477} (\bibinfo{year}{1993}).

\bibitem[{\citenamefont{J{\"u}licher et~al.}(1997)\citenamefont{J{\"u}licher,
  Ajdari, and Prost}}]{Juelicher1997}
\bibinfo{author}{\bibfnamefont{F.}~\bibnamefont{J{\"u}licher}},
  \bibinfo{author}{\bibfnamefont{A.}~\bibnamefont{Ajdari}}, \bibnamefont{and}
  \bibinfo{author}{\bibfnamefont{J.}~\bibnamefont{Prost}},
  \bibinfo{journal}{Rev. Mod. Phys.} \textbf{\bibinfo{volume}{69}},
  \bibinfo{pages}{1269} (\bibinfo{year}{1997}).

\bibitem[{\citenamefont{Reimann}(2002)}]{Reimann02}
\bibinfo{author}{\bibfnamefont{P.}~\bibnamefont{Reimann}},
  \bibinfo{journal}{Physics Reports} \textbf{\bibinfo{volume}{361}},
  \bibinfo{pages}{57} (\bibinfo{year}{2002}).

\bibitem[{\citenamefont{H\"{a}nggi and Bartussek}(1996)}]{Haenggi1996}
\bibinfo{author}{\bibfnamefont{P.}~\bibnamefont{H\"{a}nggi}} \bibnamefont{and}
  \bibinfo{author}{\bibfnamefont{R.}~\bibnamefont{Bartussek}},
  \bibinfo{journal}{Lecture Notes in Physics} \textbf{\bibinfo{volume}{476}},
  \bibinfo{pages}{294} (\bibinfo{year}{1996}).

\bibitem[{\citenamefont{H{\"a}nggi and Marchesoni}(2009)}]{Haenggi2009}
\bibinfo{author}{\bibfnamefont{P.}~\bibnamefont{H{\"a}nggi}} \bibnamefont{and}
  \bibinfo{author}{\bibfnamefont{F.}~\bibnamefont{Marchesoni}},
  \bibinfo{journal}{Rev. Mod. Phys.} \textbf{\bibinfo{volume}{81}},
  \bibinfo{pages}{387} (\bibinfo{year}{2009}).

\bibitem[{\citenamefont{Sterck et~al.}(2002)\citenamefont{Sterck, Weiss, and
  Koelle}}]{Sterck2002}
\bibinfo{author}{\bibfnamefont{A.}~\bibnamefont{Sterck}},
  \bibinfo{author}{\bibfnamefont{S.}~\bibnamefont{Weiss}}, \bibnamefont{and}
  \bibinfo{author}{\bibfnamefont{D.}~\bibnamefont{Koelle}},
  \bibinfo{journal}{Appl. Phys. A} \textbf{\bibinfo{volume}{75}},
  \bibinfo{pages}{253} (\bibinfo{year}{2002}).

\bibitem[{\citenamefont{Sterck et~al.}(2005)\citenamefont{Sterck, Kleiner, and
  Koelle}}]{Sterck2005a}
\bibinfo{author}{\bibfnamefont{A.}~\bibnamefont{Sterck}},
  \bibinfo{author}{\bibfnamefont{R.}~\bibnamefont{Kleiner}}, \bibnamefont{and}
  \bibinfo{author}{\bibfnamefont{D.}~\bibnamefont{Koelle}},
  \bibinfo{journal}{Phys. Rev. Lett.} \textbf{\bibinfo{volume}{95}},
  \bibinfo{pages}{177006} (\bibinfo{year}{2005}).

\bibitem[{\citenamefont{Sterck et~al.}(2009)\citenamefont{Sterck, Koelle, and
  Kleiner}}]{Sterck2009}
\bibinfo{author}{\bibfnamefont{A.}~\bibnamefont{Sterck}},
  \bibinfo{author}{\bibfnamefont{D.}~\bibnamefont{Koelle}}, \bibnamefont{and}
  \bibinfo{author}{\bibfnamefont{R.}~\bibnamefont{Kleiner}},
  \bibinfo{journal}{Phys. Rev. Lett.} \textbf{\bibinfo{volume}{103}},
  \bibinfo{pages}{047001} (\bibinfo{year}{2009}).

\bibitem[{\citenamefont{Weiss et~al.}(2000)\citenamefont{Weiss, Koelle, Müller,
  Gross, and Barthel}}]{Weiss2000}
\bibinfo{author}{\bibfnamefont{S.}~\bibnamefont{Weiss}},
  \bibinfo{author}{\bibfnamefont{D.}~\bibnamefont{Koelle}},
  \bibinfo{author}{\bibfnamefont{J.}~\bibnamefont{Müller}},
  \bibinfo{author}{\bibfnamefont{R.}~\bibnamefont{Gross}}, \bibnamefont{and}
  \bibinfo{author}{\bibfnamefont{K.}~\bibnamefont{Barthel}},
  \bibinfo{journal}{Europhys. Lett.} \textbf{\bibinfo{volume}{51}},
  \bibinfo{pages}{499} (\bibinfo{year}{2000}).

\bibitem[{\citenamefont{Goldobin et~al.}(2001)\citenamefont{Goldobin, Sterck,
  and Koelle}}]{Goldobin2001}
\bibinfo{author}{\bibfnamefont{E.}~\bibnamefont{Goldobin}},
  \bibinfo{author}{\bibfnamefont{A.}~\bibnamefont{Sterck}}, \bibnamefont{and}
  \bibinfo{author}{\bibfnamefont{D.}~\bibnamefont{Koelle}},
  \bibinfo{journal}{Phys. Rev. E} \textbf{\bibinfo{volume}{63}},
  \bibinfo{pages}{031111} (\bibinfo{year}{2001}).

\bibitem[{\citenamefont{Carapella}(2001)}]{Carapella2001a}
\bibinfo{author}{\bibfnamefont{G.}~\bibnamefont{Carapella}},
  \bibinfo{journal}{Phys. Rev. B} \textbf{\bibinfo{volume}{63}},
  \bibinfo{pages}{054515} (\bibinfo{year}{2001}).

\bibitem[{\citenamefont{Carapella and Costabile}(2001)}]{Carapella2001}
\bibinfo{author}{\bibfnamefont{G.}~\bibnamefont{Carapella}} \bibnamefont{and}
  \bibinfo{author}{\bibfnamefont{G.}~\bibnamefont{Costabile}},
  \bibinfo{journal}{Phys. Rev. Lett} \textbf{\bibinfo{volume}{87}},
  \bibinfo{pages}{077002} (\bibinfo{year}{2001}).

\bibitem[{\citenamefont{Carapella et~al.}(2002)\citenamefont{Carapella,
  Costabile, Martucciello, Cirillo, Latempa, Polcari, and
  Filatrella}}]{Carapella02b}
\bibinfo{author}{\bibfnamefont{G.}~\bibnamefont{Carapella}},
  \bibinfo{author}{\bibfnamefont{G.}~\bibnamefont{Costabile}},
  \bibinfo{author}{\bibfnamefont{N.}~\bibnamefont{Martucciello}},
  \bibinfo{author}{\bibfnamefont{M.}~\bibnamefont{Cirillo}},
  \bibinfo{author}{\bibfnamefont{R.}~\bibnamefont{Latempa}},
  \bibinfo{author}{\bibfnamefont{A.}~\bibnamefont{Polcari}}, \bibnamefont{and}
  \bibinfo{author}{\bibfnamefont{G.}~\bibnamefont{Filatrella}},
  \bibinfo{journal}{Physica C} \textbf{\bibinfo{volume}{382}},
  \bibinfo{pages}{337} (\bibinfo{year}{2002}).

\bibitem[{\citenamefont{Beck et~al.}(2005)\citenamefont{Beck, Goldobin,
  Neuhaus, Siegel, Kleiner, and Koelle}}]{Beck2005}
\bibinfo{author}{\bibfnamefont{M.}~\bibnamefont{Beck}},
  \bibinfo{author}{\bibfnamefont{E.}~\bibnamefont{Goldobin}},
  \bibinfo{author}{\bibfnamefont{M.}~\bibnamefont{Neuhaus}},
  \bibinfo{author}{\bibfnamefont{M.}~\bibnamefont{Siegel}},
  \bibinfo{author}{\bibfnamefont{R.}~\bibnamefont{Kleiner}}, \bibnamefont{and}
  \bibinfo{author}{\bibfnamefont{D.}~\bibnamefont{Koelle}},
  \bibinfo{journal}{Phys. Rev. Lett.} \textbf{\bibinfo{volume}{95}},
  \bibinfo{pages}{090603} (\bibinfo{year}{2005}).

\bibitem[{\citenamefont{Zapata et~al.}(1996)\citenamefont{Zapata, Bartussek,
  Sols, and H\"anggi}}]{Zapata1996}
\bibinfo{author}{\bibfnamefont{I.}~\bibnamefont{Zapata}},
  \bibinfo{author}{\bibfnamefont{R.}~\bibnamefont{Bartussek}},
  \bibinfo{author}{\bibfnamefont{F.}~\bibnamefont{Sols}}, \bibnamefont{and}
  \bibinfo{author}{\bibfnamefont{P.}~\bibnamefont{H\"anggi}},
  \bibinfo{journal}{Phys. Rev. Lett.} \textbf{\bibinfo{volume}{77}},
  \bibinfo{pages}{2292} (\bibinfo{year}{1996}).

\bibitem[{\citenamefont{Davidson et~al.}(1985)\citenamefont{Davidson, Dueholm,
  Kryger, and Pedersen}}]{Davidson1985}
\bibinfo{author}{\bibfnamefont{A.}~\bibnamefont{Davidson}},
  \bibinfo{author}{\bibfnamefont{B.}~\bibnamefont{Dueholm}},
  \bibinfo{author}{\bibfnamefont{B.}~\bibnamefont{Kryger}}, \bibnamefont{and}
  \bibinfo{author}{\bibfnamefont{N.~F.} \bibnamefont{Pedersen}},
  \bibinfo{journal}{Phys. Rev. Lett.} \textbf{\bibinfo{volume}{55}},
  \bibinfo{pages}{2059} (\bibinfo{year}{1985}).

\bibitem[{\citenamefont{Davidson et~al.}(1986)\citenamefont{Davidson, Dueholm,
  and Pedersen}}]{Davidson1986}
\bibinfo{author}{\bibfnamefont{A.}~\bibnamefont{Davidson}},
  \bibinfo{author}{\bibfnamefont{B.}~\bibnamefont{Dueholm}}, \bibnamefont{and}
  \bibinfo{author}{\bibfnamefont{N.~F.} \bibnamefont{Pedersen}},
  \bibinfo{journal}{J. Appl. Phys.} \textbf{\bibinfo{volume}{60}},
  \bibinfo{pages}{1447} (\bibinfo{year}{1986}).

\bibitem[{\citenamefont{Ustinov et~al.}(1992)\citenamefont{Ustinov, Doderer,
  Huebener, Pedersen, Mayer, and Oboznov}}]{Ustinov92}
\bibinfo{author}{\bibfnamefont{A.~V.} \bibnamefont{Ustinov}},
  \bibinfo{author}{\bibfnamefont{T.}~\bibnamefont{Doderer}},
  \bibinfo{author}{\bibfnamefont{R.~P.} \bibnamefont{Huebener}},
  \bibinfo{author}{\bibfnamefont{N.~F.} \bibnamefont{Pedersen}},
  \bibinfo{author}{\bibfnamefont{B.}~\bibnamefont{Mayer}}, \bibnamefont{and}
  \bibinfo{author}{\bibfnamefont{V.~A.} \bibnamefont{Oboznov}},
  \bibinfo{journal}{Phys. Rev. Lett.} \textbf{\bibinfo{volume}{69}},
  \bibinfo{pages}{1815} (\bibinfo{year}{1992}).

\bibitem[{\citenamefont{Martucciello and Monaco}(1996)}]{Martucciello96x}
\bibinfo{author}{\bibfnamefont{N.}~\bibnamefont{Martucciello}}
  \bibnamefont{and} \bibinfo{author}{\bibfnamefont{R.}~\bibnamefont{Monaco}},
  \bibinfo{journal}{Phys. Rev. B} \textbf{\bibinfo{volume}{53}},
  \bibinfo{pages}{3471} (\bibinfo{year}{1996}).

\bibitem[{\citenamefont{Ustinov}(2002)}]{Ustinov02a}
\bibinfo{author}{\bibfnamefont{A.~V.} \bibnamefont{Ustinov}},
  \bibinfo{journal}{Appl. Phys. Lett.} \textbf{\bibinfo{volume}{80}},
  \bibinfo{pages}{3153} (\bibinfo{year}{2002}).

\bibitem[{\citenamefont{Malomed and Ustinov}(2004)}]{Malomed04}
\bibinfo{author}{\bibfnamefont{B.~A.} \bibnamefont{Malomed}} \bibnamefont{and}
  \bibinfo{author}{\bibfnamefont{A.~V.} \bibnamefont{Ustinov}},
  \bibinfo{journal}{Phys. Rev. B} \textbf{\bibinfo{volume}{69}},
  \bibinfo{pages}{64502} (\bibinfo{year}{2004}).

\bibitem[{\citenamefont{Ustinov}(1998)}]{Ustinov98}
\bibinfo{author}{\bibfnamefont{A.~V.} \bibnamefont{Ustinov}},
  \bibinfo{journal}{Physica D} \textbf{\bibinfo{volume}{123}},
  \bibinfo{pages}{315} (\bibinfo{year}{1998}).

\bibitem[{\citenamefont{McLaughlin and Scott}(1978)}]{McLaughlin78}
\bibinfo{author}{\bibfnamefont{D.~W.} \bibnamefont{McLaughlin}}
  \bibnamefont{and} \bibinfo{author}{\bibfnamefont{A.~C.} \bibnamefont{Scott}},
  \bibinfo{journal}{Phys. Rev. A} \textbf{\bibinfo{volume}{18}},
  \bibinfo{pages}{1652} (\bibinfo{year}{1978}).

\bibitem[{\citenamefont{Borromeo et~al.}(2002)\citenamefont{Borromeo,
  Costantini, and Marchesoni}}]{Borromeo02}
\bibinfo{author}{\bibfnamefont{M.}~\bibnamefont{Borromeo}},
  \bibinfo{author}{\bibfnamefont{G.}~\bibnamefont{Costantini}},
  \bibnamefont{and}
  \bibinfo{author}{\bibfnamefont{F.}~\bibnamefont{Marchesoni}},
  \bibinfo{journal}{Phys. Rev. E} \textbf{\bibinfo{volume}{65}},
  \bibinfo{pages}{041110} (\bibinfo{year}{2002}).

\bibitem[{\citenamefont{Chauwin et~al.}(1995)\citenamefont{Chauwin, Ajdari, and
  Prost}}]{Chauwin1995:CR:AssPump}
\bibinfo{author}{\bibfnamefont{J.-F.} \bibnamefont{Chauwin}},
  \bibinfo{author}{\bibfnamefont{A.}~\bibnamefont{Ajdari}}, \bibnamefont{and}
  \bibinfo{author}{\bibfnamefont{J.}~\bibnamefont{Prost}},
  \bibinfo{journal}{Europhys. Lett.} \textbf{\bibinfo{volume}{32}},
  \bibinfo{pages}{373} (\bibinfo{year}{1995}).

\bibitem[{\citenamefont{Jung et~al.}(1996)\citenamefont{Jung, Kissner, and
  H{\"a}nggi}}]{Jung1996}
\bibinfo{author}{\bibfnamefont{P.}~\bibnamefont{Jung}},
  \bibinfo{author}{\bibfnamefont{J.~G.} \bibnamefont{Kissner}},
  \bibnamefont{and}
  \bibinfo{author}{\bibfnamefont{P.}~\bibnamefont{H{\"a}nggi}},
  \bibinfo{journal}{Phys. Rev. Lett.} \textbf{\bibinfo{volume}{76}},
  \bibinfo{pages}{3436} (\bibinfo{year}{1996}).

\bibitem[{\citenamefont{Mateos}(2000)}]{Mateos2000}
\bibinfo{author}{\bibfnamefont{J.~L.} \bibnamefont{Mateos}},
  \bibinfo{journal}{Phys. Rev. Lett.} \textbf{\bibinfo{volume}{84}},
  \bibinfo{pages}{258} (\bibinfo{year}{2000}).

\bibitem[{\citenamefont{Mateos}(2003)}]{Mateos2003:BattleOfAttractors}
\bibinfo{author}{\bibfnamefont{J.}~\bibnamefont{Mateos}},
  \bibinfo{journal}{Physica A} \textbf{\bibinfo{volume}{325}},
  \bibinfo{pages}{92} (\bibinfo{year}{2003}).

\bibitem[{\citenamefont{Son et~al.}(2003)\citenamefont{Son, Kim, Park, and
  Kim}}]{Son2004:UdmpDetRat:CurrRev}
\bibinfo{author}{\bibfnamefont{W.-S.} \bibnamefont{Son}},
  \bibinfo{author}{\bibfnamefont{I.}~\bibnamefont{Kim}},
  \bibinfo{author}{\bibfnamefont{Y.-J.} \bibnamefont{Park}}, \bibnamefont{and}
  \bibinfo{author}{\bibfnamefont{C.-M.} \bibnamefont{Kim}},
  \bibinfo{journal}{Phys. Rev. E} \textbf{\bibinfo{volume}{68}},
  \bibinfo{pages}{067201} (\bibinfo{year}{2003}).

\bibitem[{\citenamefont{Goldobin}(2003)}]{Goldobin2003}
\bibinfo{author}{\bibfnamefont{E.}~\bibnamefont{Goldobin}},
  \emph{\bibinfo{title}{Stk{J}{J} -- {U}ser's {R}eference}}
  (\bibinfo{year}{2003}),
  \urlprefix\url{http://www.geocities.com/SiliconValley/Heights/7318/StkJJ.htm}.

\end{thebibliography}

\end{document}